\documentclass[twocolumn,showpacs,preprintnumbers,amsmath,amssymb]{revtex4}

\usepackage{graphicx}
\usepackage{dcolumn}
\usepackage{bm}
\begin{document}

\title{Effective rheology of two-phase flow in a capillary fiber bundle model}
\author{Subhadeep Roy\email{subhadeep.roy@ntnu.no} and Alex
  Hansen\email{alex.hansen@ntnu.no}}

\affiliation{PoreLab, Department of Physics, Norwegian University of
  Science and Technology, NO--7491 Trondheim, Norway}

\author{Santanu Sinha\email{santanu@csrc.ac.cn}}

\affiliation{Beijing Computational Sciences Research Center, 10 East
  Xibeiwang Road, Haidian District, Beijing 100193, China}
\date{\today {}}
\begin{abstract}
We investigate the effective rheology of two-phase flow in a bundle of
parallel capillary tubes carrying two immiscible fluids under an
external pressure drop. The diameter of the tubes vary along the
length which introduce capillary threshold pressures. We demonstrate
through analytical calculations that a transition from a linear Darcy
to a non-linear behavior occurs while decreasing the pressure drop
$\Delta P$, where the total flow rate $\langle Q \rangle$ varies with
$\Delta P$ with an exponent $2$ as $\langle Q \rangle\sim \Delta P^2$
for uniform threshold distribution. The exponent changes when a lower
cut-off $P_m$ is introduced in the threshold distribution and in the
limit where $\Delta P$ approaches $P_m$, the flow rate scales as
$\langle Q \rangle \sim (|\Delta P|-P_m)^{3/2}$. While considering
threshold distribution with a power $\alpha$, we find that the
exponent $\gamma$ for the non-linear regime vary as $\gamma=\alpha+1$
for $P_m=0$ and $\gamma=\alpha+1/2$ for $P_m>0$. We provide numerical
results in support of our analytical findings.       
\end{abstract}
\maketitle

Understanding the hydrodynamic properties of simultaneous flow of two
or more immiscible fluids is essential due its relevance to a wide
variety of different systems in industrial, geophysical and medical
sectors \cite{b72,d92}. Different applications, such as bubble
generation in microfluidics, blood flow in capillary vessels, catalyst
supports used in the automotive industry, transport in fuel cells, oil
recovery, ground water management and ${\rm CO}_2$ sequestration, deal
with the flow of bubble trains in different types of systems, ranging
from single capillaries to more complex porous media. The underlying
physical mechanisms in multiphase flow are controlled by a number of
factors, such as the capillary forces at the interfaces, viscosity
contrast between the fluids, wettability and geometry of the system,
which make the flow properties different compared to single phase
flow. When one immiscible fluid invades a porous medium filled with
another fluid, different types of transient flow patterns, namely
viscous fingering \cite{cw85, mfj85}, stable displacement \cite{ltz88}
and capillary fingering \cite{lz85} are observed while tuning the
physical parameters \cite{lmtsm04}. These transient flow patterns were
modeled by invasion percolation \cite{ww83} and diffusion limited
aggregation (DLA) models \cite{ws81}. When steady state sets in after
the initial instabilities, the flow properties in are characterized by
relations between the global quantities such as flow rate, pressure
drop and fluid saturation \cite{v18,hsbkgv18}. It has been observed
theoretically and experimentally that, in the regime where capillary
forces compete with the viscous forces, the two-phase flow rate of
Newtonian fluids in the steady state no longer obeys the linear Darcy
law \cite{d56,w86} but varies as a power law with the applied pressure
drop \cite{tkrlmtf09,tlkrfm09,rcs11,sbdk17}. Tallakstad et
al.\ \cite{tkrlmtf09,tlkrfm09} experimentally measured the exponent of
the power law to be close to two ($=1/0.54$) in a two-dimensional
system and followed this observation up with arguments why the
exponent should be two. Rassi et al.\ \cite{rcs11} found a value for
the exponent varying between $2.2$ ($=1/0.45$) and $3.0$ ($=1/0.33$)
in a three-dimensional system. Sinha et al.\ \cite{sbdk17} considered
a similar system to that which had been studied by Rassi et
al.\ finding an exponent $2.17\pm 0.24$ ($=1/(0.46\pm 0.05)$). The
reason behind the discrepancy between the results of Rassi et al.\ and
those of Sinha et al.\ is the possibility of a non-zero threshold
pressure that observed in the later study, under which there would be
no flow, which was assumed to be zero in the former study. The
reciprocals in the brackets are provided in order to compare the
exponent values reported in the literature \cite{tkrlmtf09, tlkrfm09,
  rcs11, sbdk17} with those we present here in this article, as we
express our results as $\langle Q \rangle$ as a power law in $\Delta P$,
whereas the cited papers write $\Delta P$ as a power law in $\langle Q \rangle$.
 
This power law behavior is in contrast to the assumption of linearity
in the relation between flow rate and pressure drop that is generally
assumed in the relative permeability approach dominating reservoir
simulations \cite{wb36}.

For a single capillary tube with varying diameter, Sinha et
al.\ \cite{shbk13} showed that the average volumetric flow rate
$\overline{q}$ in the steady state has a non-linear square-root type
relationship with the pressure drop $\Delta P$ as $\overline{q}\sim
\sqrt{\Delta P^2 - P_c^2}$. This was shown analytically by integrating
the instantaneous linear two-phase flow equation over the whole
capillary tube. Here $P_c$ is the threshold pressure difference below
which there is no flow.  It appears due to the capillary barriers at
the interfaces at the narrow pore throats. Extending this non-linear
relationship to a network of disordered pores, the relationship
between the steady-state flow rate and an excess pressure drop leads
to a quadratic relationship in the capillary dominated regime
\cite{sh12}.  The quadratic relationship for the pore network, both in
two and in three dimensions, was obtained analytically by mean-field
calculations and numerically with pore network modeling
\cite{sbdk17,sh12}.

While increasing the pressure drop, the capillary forces become
negligible compared to the viscous forces. This leads to a crossover
from the non-linear regime to a linear Darcy regime for both the
single capillary and for the pore network. Such non-linear quadratic
relationship at low flow rate and a crossover to a linear regime at
high flow rate was also observed in case of the single-phase flow of
Bingham viscoplastic fluid in porous media \cite{rh87, ct15}. A
Bingham fluid is a yield threshold fluid which behaves like a solid
below the threshold and flows like a Newtonian fluid above it. The
origin of the quadratic regime for the Bingham fluid flowing in a
porous media can be understood intuitively in this way: the flow
starts when one connected channel appears in the system just above a
threshold pressure and the flow rate varies linearly with the excess
pressure drop; while increasing the applied pressure drop further,
more number of connected flow channels start to appear enhancing the
overall flow rate more rapidly than the applied pressure drop leading
to the quadratic relationship. Finally, when all possible flow paths
become active, the flow become Newtonian following the linear
Darcy law. Note that, in general, the rheology of the Bingham fluid is
linear above the yield threshold.  It is the disorder in the yield thresholds
due to the porous medium that creates the quadratic regime.

The argument presented by Tallakstad et al.\ \cite{tkrlmtf09,tlkrfm09}
focused on the successive opening of fluid channels when the pressure
drop across the system was increased. When $|\Delta P|$ is small, the
flow will occur along isolated channels. The volumetric flow rate in
such a channel will be proportional to $|\Delta P|/L$.  Between the
channels there will be fluid clusters held in place by capillary
forces, say of the order $p_t$. There is a pressure gradient $|\Delta
P|/L$ in the flow direction. A given cluster of length $l_{\parallel}$
will be stuck if $p_t > l_{\parallel} |\Delta P|/L$. The largest stuck
cluster will then have a size $l_{\parallel,\max}=Lp_t/|\Delta P|$. If
we now assume that this length, $l_{\parallel,\max}$ is same as the
distance between the channels where there is flow, $l_{\perp}$, then
the total flow rate must be equal to the number of channels, which is
proportional to $1/l_{\perp}$, multiplied by the flow rate in each
channel. Hence, we have $Q \propto (1/l_{\perp})\ |\Delta P| \propto
|\Delta P|^2$. Though this argument provides the same behavior as the
one based on the mean field calculation \cite{sh12} for
two-dimensional networks, a difference appears in three
dimensions. Following the same arguments, it leads to the flow rate
varying as the pressure drop to the {\it third\/} power as long as the
isolated channels remain one-dimensional strings rather than
two-dimensional sheets in three dimensions.

We present in this article a capillary fiber bundle model
\cite{s53,s74}, which is a system of $N$ parallel capillary tubes,
disconnected from each other, each carrying an independent bubble
trail of two immiscible fluids under an external pressure drop. In a
porous medium, a typical pore constitutes of two wide pore bodies at
the ends and a narrow pore throat in the middle. When an interface
moves along the pore, the capillary pressure associated with the
interface becomes position dependent due to the change in the radius of
curvature. This introduces an overall threshold pressure that depends
on the position of all the interfaces \cite{shbk13}. One can simplify
such shape of the pore by a sinusoidal type and a long capillary tube
with varying radius can be seen as a series of many pores. In the
capillary bundle model, the diameter of each tube varies along the
axis identically and the disorder in the threshold appear due to the
different interface positions in different tubes. This model is
essentially the only model for immiscible two-phase flow which is
analytically tractable. We calculate the total average flow rate as a
function of the applied pressure drop and study the effect of disorder
in the threshold distribution. We point out that, here we do not
address the question of the relation between the fluid
distributions in the capillaries and the respective threshold
distributions. Our aim with this model is to investigate how the range of the
disorder in the threshold distribution controls the effective flow
properties. This provides insight into the
non-linearities in steady-state two-phase flow. We will see that the
exponent for the non-linear regime depends on the lower cut-off of the
threshold distribution as well as on the behavior of the distribution
near the cut-off. The possibility to study analytically for this model
how the competition between viscous and capillary forces renders the
Darcy relation non-linear, is a new and useful discovery.

\begin{figure}
\centerline{\includegraphics[width=0.5\textwidth,clip]{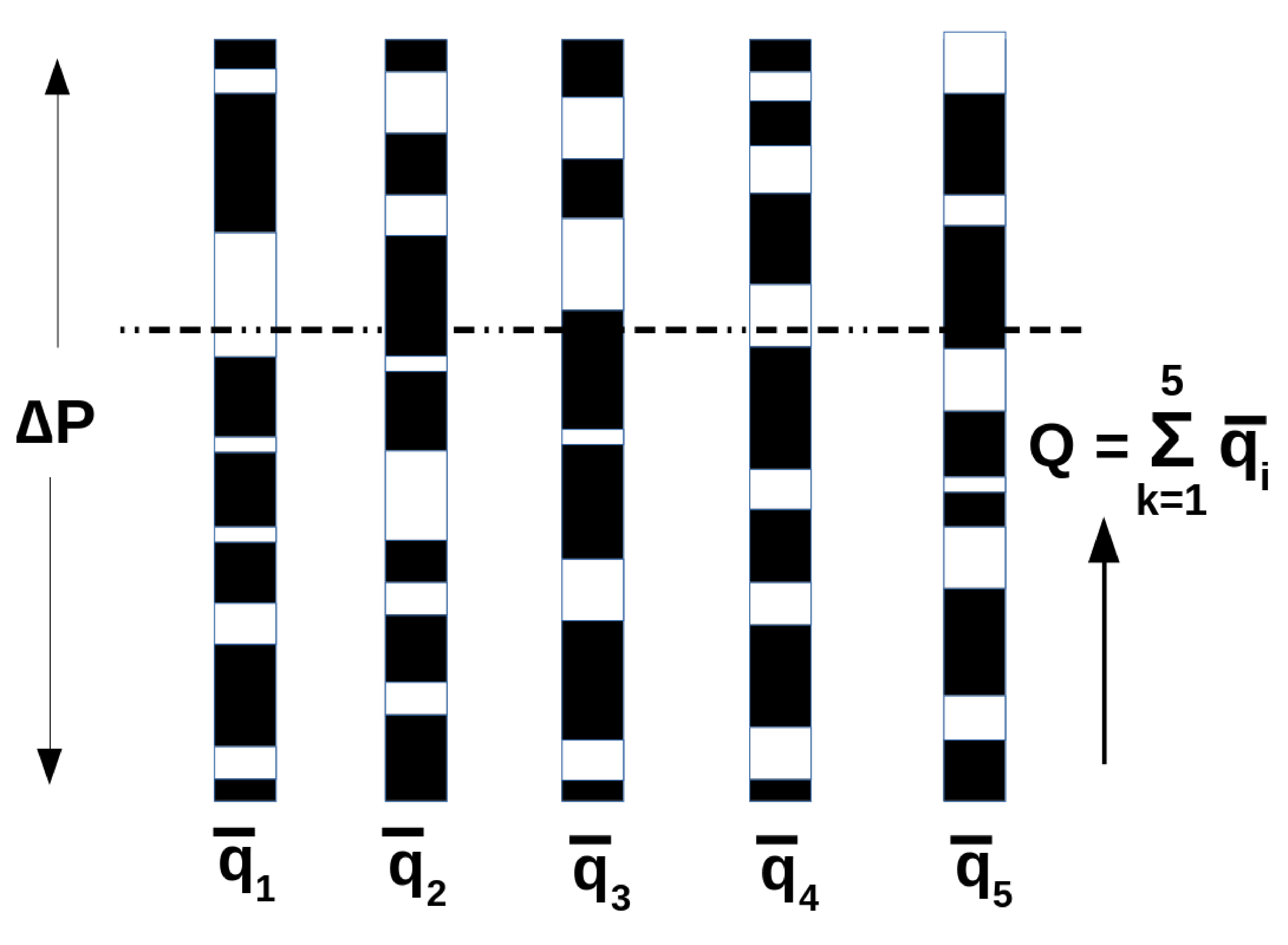}}
\caption{The capillary tube model. There are $N=5$ capillaries, each
  filled with a bubble train of wetting (white) and non-wetting
  (black) fluids moving in the direction of the arrow. The diameter
  along each tube vary so that the capillary force from each interface
  vary with its position. The varying diameters are not illustrated in
  the figure. The average diameters are the same for all tubes. The
  broken line illustrates an imaginary cut through the capillary fiber
  bundle.}
\label{fig1}
\end{figure}

The capillary fiber bundle model is a hydrodynamic analog of the fiber
bundle model used in fracture mechanics to model mechanical failure
under stress \cite{hhp15}. The fiber bundle model is an ideal example
of a disordered system in statistical mechanics driven by threshold
activated dynamics. It is a simple, yet very rich model to understand
failure events in mechanical systems. In its simplest form it is
analytically tractable. In more complex versions of the model,
analytical calculations go hand in hand with numerical simulations.

Figure \ref{fig1} illustrates a bundle with $N=5$ parallel
capillaries. Each capillary tube has a length $L$ and an average inner
area $a$. For each capillary, the diameter varies along the long axis
identically. Each capillary is filled with a bubble train of wetting
and non-wetting fluids. Due to the varying diameter, the capillary
forces at the interfaces vary as the bubble train moves along the
tubes. We assume that the wetting fluid does not form films along the
pore walls so that the fluids do not pass each other. The lengths of
the wetting and non-wetting fluids in each tube is $L_w$ and $L_n$
respectively such that the volume of the wetting fluid in each tube is
$L_w a$ and the volume of the non-wetting fluid is $L_n a$, where $a$
is the average cross-sectional area of the capillary tubes. Hence, the
saturations are $S_w=L_w/L$ and $S_n=L_n/L$ for each capillary
tube. The cross-sectional pore area of the capillary fiber bundle is
\begin{equation}
\label{eqn23-0}
A_p=Na\;.
\end{equation} 

Though each tube contains the same amount of each fluid, it has its
own division of the fluids into bubbles. We average over the ensemble
of capillary tubes which constitute the capillary fiber bundle by
averaging over the fluids in each tube that passes at a given instance
through the imaginary cut shown in the figure. We will obtain the same
averages if we consider a single capillary tube, averaging over a time
interval the fluid passing the imaginary cut
\cite{shbk13,sshb16}. This shows that the model is ergodic.

The volumetric flow rate in a capillary tube is given by \cite{shbk13}
\begin{equation}
\label{eqn1}
q=-\frac{a^2}{8\pi\mu_{\rm av} L}\Theta(|\Delta P|-P_c)\left[\Delta P-P_c\right]\;,
\end{equation}
where $\Delta P$ is the pressure drop across the capillary tube, $P_c$
the sum of all the capillary forces along the capillary tube due to
the interfaces and
\begin{equation}
\label{eqm2}
\mu_{\rm av}=S_w\mu_w+S_n\mu_n\;,
\end{equation}
is the effective viscosity. $\Theta(|\Delta P|-P_c)$ is the Heaviside
function which is zero for negative arguments and one for positive
arguments. The function ${\rm sgn }(\Delta P)$ is the sign of the
argument.

Sinha et al.\ \cite{shbk13} showed that the time average when the
pressure difference across the tube is kept fixed is given by
\begin{eqnarray}
\label{eqn3}
&&\overline{q}(P_c)=\nonumber\\
&&-\frac{a^2}{8\pi\mu_{\rm av} L}{\rm sgn}(\Delta P)\Theta(|\Delta P|-P_c)\sqrt{\Delta P^2-P_c^2}\;, 
\end{eqnarray}

Suppose now that the thresholds $P_c$ are distributed uniformly
between zero and a maximum value $P_M$.  The cumulative threshold
probability is then
\begin{equation}
\label{eqn4}
\Pi(P_c)=\left\{\begin{array}{ll}
                                0       & \mbox{, $P_c \le 0$\;,}\\
                                \frac{P_c}{P_M} & \mbox{, $0 < P_c \le P_M$\:,}\\
                                1       & \mbox{, $P_c > P_M$\;.}\\
              \end{array}
       \right.
\end{equation}

We have $N$ capillary tubes. Using order statistics, we may order the
$N$ averaged threshold values,
\begin{equation}
\label{eqn5}
\Pi(P_c(k))=\frac{k}{N+1}\;,
\end{equation}
where $1 \le k\le N$. Hence,
\begin{equation}
\label{eqn6}
P_c(k)=P_M\frac{k}{N+1}\;.
\end{equation}

The average volumetric flow rate through the capillary fiber bundle
for $|\Delta P|>0$ is then
\begin{equation}
\label{eqn7}
\langle Q\rangle = \sum_{k=1}^{(N+1)\min\left(\frac{|\Delta
    P|}{P_M},1\right)} \overline{q}\left(P_c(k)\right)
\end{equation}
We assume the limit $N\to\infty$ turning the sum into an integral,
\begin{eqnarray}
\label{eqn8}
\frac{\langle Q\rangle}{N}&=&-\frac{a^2 P_M{\rm sgn}(\Delta
  P)}{8\pi\mu_{\rm av} L}\nonumber\\ &&\int_0^{\min(|\Delta P|/P_M,1)}
dx\sqrt{\left(\frac{|\Delta P|}{P_M}\right)^2-x^2}\;.
\end{eqnarray}
This integral is doable and we find
\begin{equation}
\label{eqn9}
\frac{\langle Q\rangle}{N}=-\frac{a^2}{32\mu_{\rm
    av}L}\left|\frac{\Delta P}{P_M}\right|\Delta P
\end{equation}
when $|\Delta P| \le P_M$ and
\begin{eqnarray}
\label{eqn9-1}
&&\frac{\langle Q\rangle}{N}=-\frac{a^2 P_M{\rm sgn}(\Delta
  P)}{16\pi\mu_{\rm av}
  L}\nonumber\\ &&\left[\sqrt{\left(\frac{|\Delta
      P|}{P_M}\right)^2-1}\right.\nonumber\\ &&\left.+\left(\frac{|\Delta
    P|}{P_M}\right)^2{\rm arcsin}\left(\frac{P_M}{|\Delta P|}\right)
  \right]\;,
\end{eqnarray}
when $|\Delta P| > P_M$. In the limit $|\Delta P| \gg P_M$, Equation
(\ref{eqn9-1}) gives
\begin{equation} 
\label{eqn9-3}
\frac{\langle Q\rangle}{N}=-\frac{a^2 }{8\pi\mu_{\rm av} L}\ \Delta
P\;.
\end{equation}
Hence, the Darcy relation for a tube is recovered.

We see that this picture is consistent with that central to the
arguments of Tallakstad et al.\ \cite{tkrlmtf09,tlkrfm09} leading to
the quadratic dependence of $Q$ on $\Delta P$. From Equation
(\ref{eqn6}) we deduce that a number $k_c$ of the capillary tubes are
active, where
\begin{equation}
\label{eqn12}
k_c=\frac{|\Delta P|}{P_M}\ (N+1)\;.
\end{equation}
The typical distance between active capillary tubes in units of the
distance between the tubes is then given by
\begin{equation}
\label{eqn13}
l_{\perp}=\frac{N+1}{k_c}=\frac{P_M}{|\Delta P|}\;,
\end{equation}
in accordance with the argument of Tallakstad et al.

How stable is the square law $Q \propto |\Delta P|^2$? That is, how
much does it hinge on the choice of cumulative threshold probability
$\Pi(P_c)$. So far we have only considered the one given in Equation
(\ref{eqn4}). Let us now generalize it to
\begin{equation}
\label{eqn4-1}
\Pi(P_c)=\left\{\begin{array}{ll} 0 & \mbox{, $P_c \le
  0$\;,}\\ \left(\frac{P_c}{P_M}\right)^\alpha & \mbox{, $0 < P_c \le
  P_M$\:,}\\ 1 & \mbox{, $P_c > P_M$\;,}\\
              \end{array}
       \right.
\end{equation}
where $\alpha > 0$.  The average ordered threshold are then given by
\begin{equation}
\label{eqn6-1}
P_c(k)=P_M\left(\frac{k}{N+1}\right)^{1/\alpha}\;,
\end{equation}
and when combined with the expression for $\langle Q\rangle$, Equation
(\ref{eqn7}) in the limit $N\to\infty$, we find
\begin{eqnarray}
\label{eqn8-1}
\frac{\langle Q\rangle}{N}&=&-\frac{a^2 P_M{\rm sgn}(\Delta
  P)}{8\pi\mu_{\rm av} L}\nonumber\\ &&\int_0^{\min\left((|\Delta
  P|/P_M)^\alpha,1\right)} dx\sqrt{\left(\frac{|\Delta
    P|}{P_M}\right)^2-x^{2/\alpha}}\;.\nonumber\\
\end{eqnarray}
Since we are interested in the behavior for $|\Delta P| \to 0$, we do
this integral under the assumption that $|\Delta P| < P_M$ finding
\begin{equation}
\label{eqn8-2}
\frac{\langle Q\rangle}{N}=-\frac{a^2\alpha}{32\sqrt{\pi}\mu_{\rm av} L}
\frac{\Gamma\left(\frac{\alpha}{2}\right)}{\Gamma\left(\frac{3+\alpha}{2}\right)}
\left(\frac{|\Delta P|}{P_M}\right)^\alpha\ \Delta P\;,
\end{equation}
where the $\Gamma$ function for real positive $z$ is defined as,
$\Gamma (z)=
\displaystyle\int_{-\infty}^{\infty}t^{z-1}e^{-t}dt$. When $\alpha=1$,
we recover Equation (\ref{eqn9}).

Equation (\ref{eqn9}) shows the behavior observed experimentally in References
\cite{tkrlmtf09} and \cite{tlkrfm09}.  With Equation (\ref{eqn8-2}), we have just
shown that $\langle Q\rangle/N \sim |\Delta P|^\gamma$ as $|\Delta P| \to 0$, 
where $\gamma$ depends on the
threshold distribution, i.e.\ on $\alpha$ in Equation (\ref{eqn4-1}).  Does this imply that
there is no universality; that the experimentally observed behavior is due to the
presence of a very specific threshold distribution?  

As we now argue, there is universality. We note that the threshold distribution
$p(P_c)=d\Pi(P_c)/dP_c$ behaves as $p(P_c)\propto P_c^{\alpha-1}$.
Hence, if $\alpha >1$, the distribution vanishes as $P_c\to 0$,
whereas it diverges for $\alpha < 1$.  Hence, the behavior of the
distribution is vastly different for these two cases, and this causes $\gamma$ to 
depend on $\alpha$.  However, for $\alpha=1$, the distribution reaches a constant, 
non-zero value for $P_c\to 0$.      
Any threshold distribution with this behavior for small $P_c$, i.e., $p(P_c)$
reaching a non-zero value and $dp(P_c)/dP_c\to 0$ in the limit $P_c\to 0$ will give
rise to the square power law seen in Equation (\ref{eqn9}).  Such distributions are
ubiquitous, and $\gamma=2$ is universal over this class of distributions. 
      
We now consider $\alpha=1$ again, but introduce a minimum threshold
$P_m$ so that the cumulative threshold probability is given by
\begin{equation}
\label{eqn10}
\Pi(P_c)=\left\{\begin{array}{ll}
                                0       & \mbox{, $P_c \le P_m$\;,}\\
                                \frac{P_c-P_m}{P_M-P_m} & \mbox{, $P_m < P_c \le P_M$\:,}\\
                                1       & \mbox{, $P_c > P_M$\;.}\\
              \end{array}
       \right.
\end{equation}
Equation (\ref{eqn5}) yields in this case the ordered threshold sequence
\begin{equation}
\label{eqn14}
P_c(k)=P_m+(P_M-P_m)\frac{k}{N+1}\;.
\end{equation}
Equation (\ref{eqn7}) now becomes in the limit $N\to\infty$
\begin{eqnarray}
\label{eqn15}
\frac{\langle Q\rangle}{N}&=&-\frac{a^2 (P_M-P_m){\rm sgn}(\Delta P)}{8\pi\mu_{\rm av} L}\nonumber\\
&&\int_{\frac{P_m}{P_M-P_m}}^{\frac{|\Delta P|}{P_M-P_m}} dx
\sqrt{\left(\frac{|\Delta P|}{P_M-P_m}\right)^2-x^2}\;,\nonumber\\
\end{eqnarray}
when we assume $P_m \le |\Delta P| \le P_M$.  We find
\begin{eqnarray}
\label{eqn16}
\frac{\langle Q\rangle}{N}&=&-\frac{a^2 (P_M-P_m){\rm sgn}(\Delta P)}{32\pi\mu_{\rm av} L}\left(\frac{|\Delta P|}{P_M-P_m}\right)^2\nonumber\\
&&\left[\pi-4\left(\frac{P_m}{|\Delta P|}\right)^2\sqrt{\left(\frac{|\Delta P|}{P_m}\right)^2-1}\right.\nonumber\\ 
&-&\left.2{\rm arccot}\left(\frac{2\sqrt{\left(\frac{|\Delta P|}{P_m}\right)^2-1}}{2-\left(\frac{|\Delta P|}{P_m}\right)^2}\right)\right]\;.\nonumber\\
\end{eqnarray}
We find to lowest order in $(|\Delta P|-P_m)$, that this expression behaves as 
\begin{equation}
\label{eqn17}
\frac{\langle Q\rangle}{N}=-\frac{a^2{\rm sgn}(\Delta P)}{3\sqrt{2}\pi\mu_{\rm av} L}\frac{\sqrt{P_m}}{(P_M-P_m)}(|\Delta P|-P_m)^{3/2}\;,
\end{equation}
as $|\Delta P|\to P_m$. \\

\begin{figure*}
  \centerline{\hfill
    \includegraphics[width=0.45\textwidth,clip]{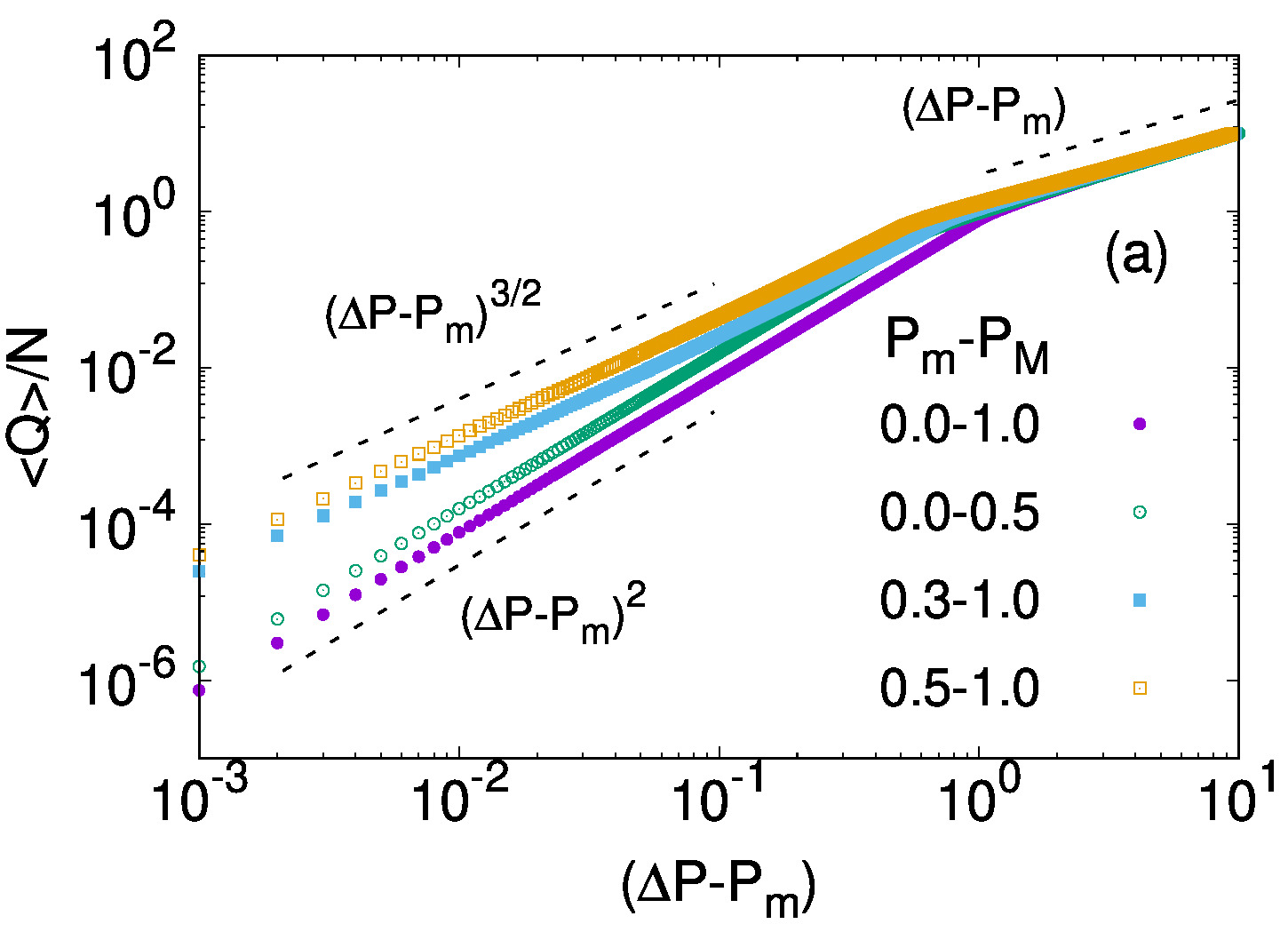}\hfill
    \includegraphics[width=0.45\textwidth,clip]{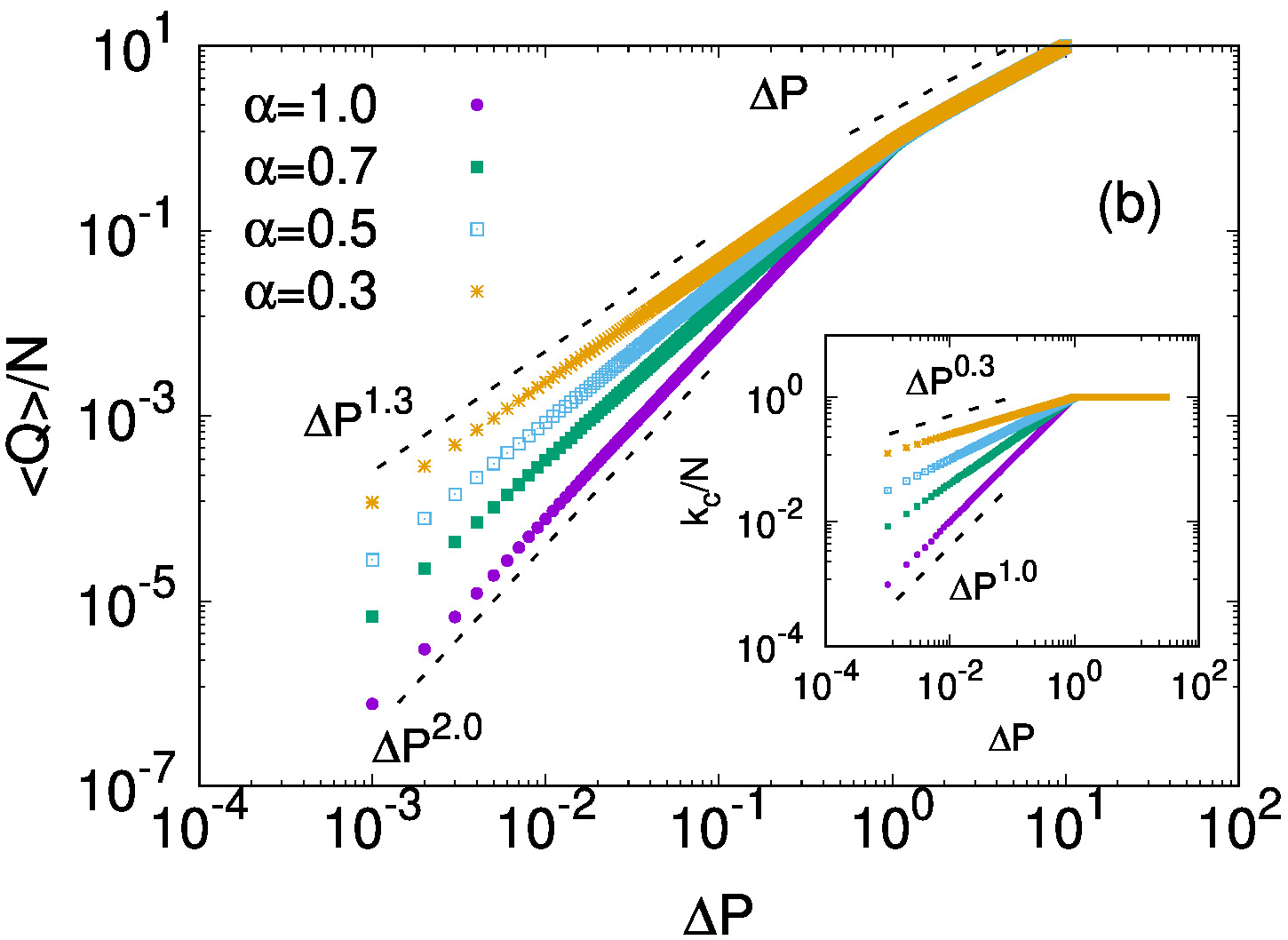}\hfill
  }
  \centerline{\hfill
    \includegraphics[width=0.45\textwidth,clip]{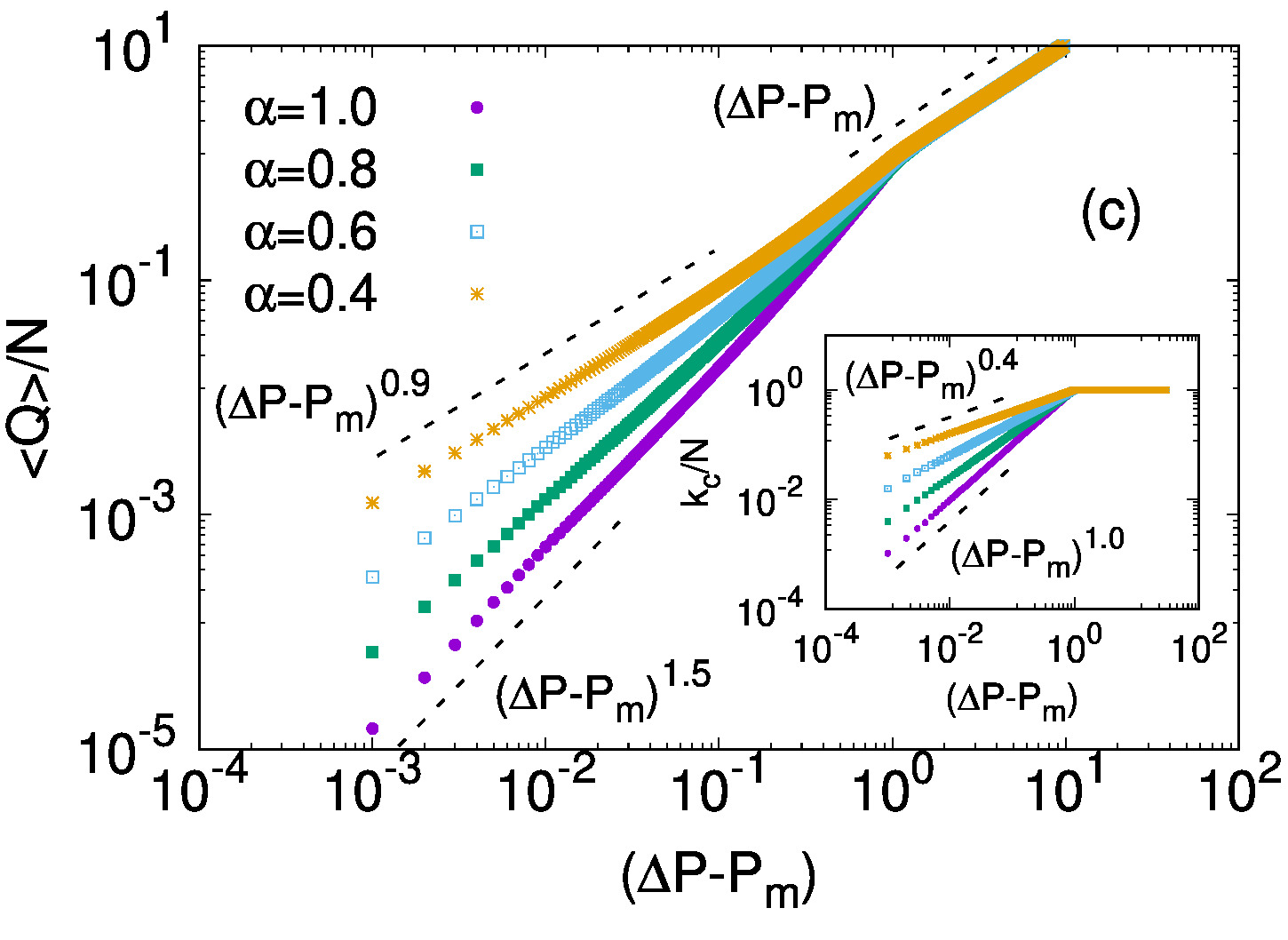}\hfill
    \includegraphics[width=0.45\textwidth,clip]{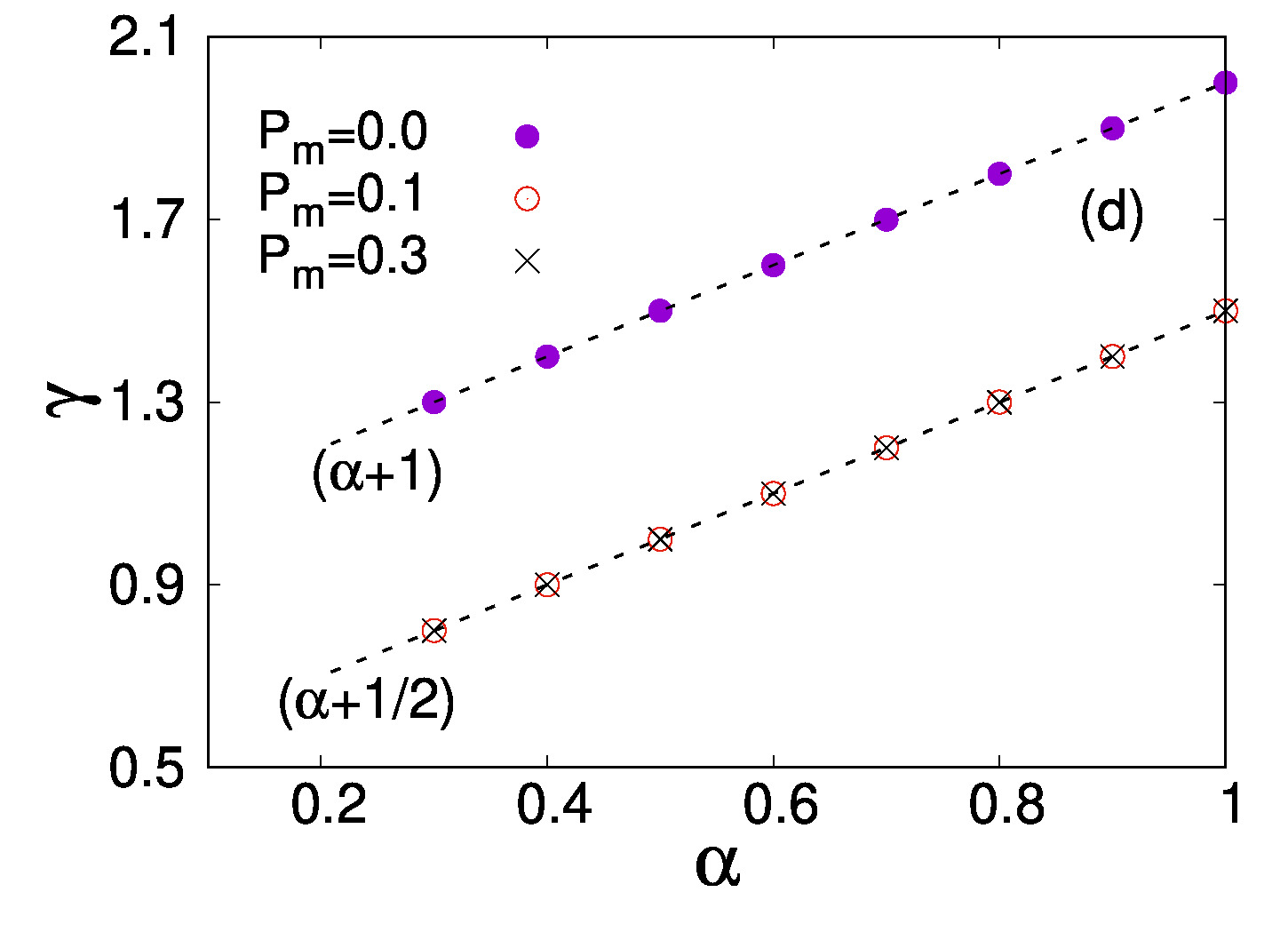}\hfill
  }
\caption{\label{fig2} Results from numerical simulations performed
  with $N=10^5$ capillaries and averaged over $10^4$
  configurations. Variation of $\langle Q \rangle/N$ as a function of
  pressure drop $\Delta P$ for different threshold distributions are
  shown in (a), (b) and (c) where non-linear to linear transitions are
  observed while increasing the pressure drop. Figure (a) corresponds
  to uniform threshold distribution [Equation (\ref{eqn4})] where the
  power-law exponent $\gamma$ for the non-linear regime has a value
  $2$ without a lower cut-off ($P_m=0$). With any non-zero lower
  cut-off ($P_m>0$), the exponent shifts to $3/2$ [Equation
    (\ref{eqn17})]. Results for the threshold distribution with a
  power $\alpha$ [Equation (\ref{eqn4-10})] are shown in (b) and (c)
  for $P_m=0$ and $P_m>0$ respectively, where $\gamma$ varies with
  $\alpha$ as $\gamma=\alpha +1$ for $P_m=0$ and as
  $\gamma=\alpha+1/2$ for $P_m>0$. These two relations are shown in
  (d) for the range of $\alpha$ which show two distinct straight lines
  for $P_m=0$ and for $P_m>0$. Here, the number of active capillaries
  ($k_c$) vary with $\Delta P$ as $k_c/N\sim(\Delta P-P_m)^\alpha$ as
  shown in the insets of (b) and (c).}
\end{figure*}

We now turn to numerical simulations and observe that the numerical
results are in good a agreement with the analytical findings. The
numerical simulations also allow us to explore the regions which are
analytically challenging. Results are shown in Figure \ref{fig2} for a
bundle containing $N=10^5$ capillary tubes and averaged over $10^4$
configurations. In Figure \ref{fig2}(a), we show the behavior of the
volumetric flow rate $\langle Q \rangle$ as a function of increasing
pressure drop $\Delta P$ for uniform threshold distributions with
$P_m=0$ and $P_m>0$, given by Equations (\ref{eqn4}) and (\ref{eqn10})
respectively. The results show that, for each threshold distribution,
the relationship is linear for high $\Delta P$ obeying the Darcy law
as predicted by Equation (\ref{eqn9-3}). For small pressure drops,
$\langle Q \rangle$ follows a power law in $\Delta P$ with an exponent
$2$ when there is no lower cut-off in the threshold distribution,
i.e. $P_m=0$. This is predicted in Equation (\ref{eqn9}). When a lower
cut-off is introduced in the threshold distribution ($P_m>0$), this
exponent shifts from $2$ to $3/2$ as predicted in Equation
(\ref{eqn17}). These exponents in the non-linear regime are not
sensitive to the span of the distribution as shown in Figure
\ref{fig2} (a). An insight to a more generalized picture is presented
in Figure \ref{fig2}(b) and (c) for a threshold distribution given by
a generalization of Equation (\ref{eqn4-1}) with an introduction of a
lower cut-off $P_m$,
\begin{equation}
\label{eqn4-10}
\Pi(P_c)=\left\{\begin{array}{ll}
                                0       & \mbox{, $P_c \le P_m$\;,}\\
                                \left(\frac{P_c-P_m}{P_M-P_m}\right)^\alpha & \mbox{, $P_m < P_c \le P_M$\:,}\\
                                1       & \mbox{, $P_c > P_M$\;,}\\
              \end{array}
       \right.
\end{equation}
With this distribution of thresholds, the exponent $\gamma$ in the
non-linear region shows a continuous variation with $\alpha$ as
$\gamma=\alpha+1$ for $P_m=0$. Such variation is given in Equation
(\ref{eqn8-2}) and matches well with the numerical results as shown in
see Figure \ref{fig2} (d). In presence of a lower cut-off $P_m>0$,
$\gamma$ varies as $(\alpha+1/2)$ instead of $(\alpha+1)$ irrespective
of the position of the lower cut-off. An analytical treatment for a
general $\alpha$ value with $P_m>0$ is rather
challenging. Nevertheless, our numerical result matches with the
analytical study (see Equation \ref{eqn17}) in the limit $\alpha=1$.

Equation (\ref{eqn8-2}) predicts an exponent $\gamma=\alpha+1$. A simple argument,
related to that given by Roux and Herrmann \cite{rh87}, goes as
follows: The number of active capillary tubes is proportional to
$(|\Delta P|-P_m)^\alpha$. This behavior is observed in the insets in
Figures \ref{fig2} (b) and (c). The flow rate in an active capillary
is proportional to $(|\Delta P|-P_m)^{1/2}$. Hence, the total flow
rate should be $\langle Q\rangle \propto (|\Delta P|-P_m)^{\alpha+1/2}$. It is
accidental that this argument works out for $P_m>0$ (Figure \ref{fig2}
(c)), as it does {\it not\/} when $P_m=0$, where $\gamma=\alpha+1$. For the argument
to function, the distribution of active capillaries and the flow rate in each 
capillary should be uncorrelated.  They are not. 

We find the same behavior with respect to the cut-off: An exponent 3/2
for the cumulative threshold probability
\begin{equation}
\label{eqn4-11}
\Pi(P_c)=\left\{\begin{array}{ll}
                                0       & \mbox{, $P_c \le P_m$\;,}\\
                                \frac{\log\left(\frac{P_c}{P_m}\right)}{\log\left(\frac{P_c}{P_M}\right)} & \mbox{, $P_m < P_c \le P_M$\:,}\\
                                1       & \mbox{, $P_c > P_M$\;,}\\
              \end{array}
       \right.
\end{equation}
where $P_m=10^{-\beta}$ and $P_M=10^\beta$ and $\beta$ ranging from
$0.5$ to $1.5$. The same goes for the cumulative threshold probability
\begin{equation}
\label{eqn4-12}
\Pi(P_c)=\left\{\begin{array}{ll}
                                0       & \mbox{, $P_c \le P_m$\;,}\\
                                1-e^{-P_c-P_m/P_d} & \mbox{, $P_m < P_c$\:,}\\
              \end{array}
       \right.
\end{equation}
where we have set $P_m=0.1$ and $P_d=1$. In both of these cases, the
probability density at $P_c=P_m$ is finite.

We have presented an analytical study supported by numerical
simulations of steady-state two-phase flow in a system of parallel
capillary tubes.  Considering a uniform distribution for the threshold
pressures for the capillaries, we have calculated the average flow
rate as a function of the applied pressure drop. When the thresholds
are distributed according to a uniform distribution between zero and a
maximum value --- or more generally, the threshold distribution
approaches a non-zero value in the limit of zero thresholds --- we
obtain a quadratic relationship between the flow rate and the applied
pressure drop when the applied pressure drop is below the maximum
threshold pressure, and the linear Darcy relationship for higher
pressure drops. This crossover between a quadratic non-linear and
linear flow regimes is in agreement with many existing results of
two-phase flow in porous media which shows that this simple model can
capture effective two-phase flow properties of more complex porous
media. When a lower cut-off is introduced in the threshold
distribution, the quadratic relationship changes, and the flow rate
varies with an excess pressure drop with an exponent $3/2$ as the
pressure drop approaches to the lowest threshold pressure.

The difference between the capillary fiber bundle model and a porous
medium is that in the latter, the fluids meet and mix at the nodes of
the pore network.  This is an essential mechanism that leads to the
non-linear Darcy law is a power law with an exponent two as seen in
the experiments, the numerical simulations and the mean-field
calculations.  However, it remains a mystery how the mixing at the
nodes leads to this universality.

\bigskip

The authors thank Dick Bedeaux, Carl Fredrik Berg, Eirik
G.\ Flekk{\o}y, Signe Kjelstrup, Knut J{\o}rgen M{\aa}l{\o}y, Per Arne
Slotte and Ole Tors{\ae}ter for interesting discussions. This work was
partly supported by the Research Council of Norway through its Centres
of Excellence funding scheme, project number 262644. SS was supported
by the National Natural Science Foundation of China under grant number
11750110430.


\begin{thebibliography}{99}
\bibitem{b72} Bear J. {\it Dynamics of fluids in porous
  media}. Mineola, NY: Dover (1988).

\bibitem{d92} Dullien FAL. {\it Porous media: fluid, transport and
  pore structure}. San Diego: Academic Press (1992).

\bibitem{cw85} Chen JD, Wilkinson D. Pore-scale viscous fingering in
  porous media. {\it Phys Rev Lett}. (1985) {\bf 55}:1892. doi:
  10.1103/PhysRevLett.55.1892

\bibitem{mfj85} M{\aa}l{\o}y KJ, Feder J, J{\o}ssang T. Viscous
  fingering fractals in porous media. {\it Phys Rev Lett}. (1985) {\bf
    55}:2688. doi: 10.1103/PhysRevLett.55.2688

\bibitem{ltz88} Lenormand R, Touboul E, Zarcone C. Numerical models
  and experiments on immiscible displacements in porous media. {\it J
    Fluid Mech}. (1988) {\bf 189}:165. doi: 10.1017/S0022112088000953

\bibitem{lz85} Lenormand R, Zarcone C. Invasion percolation in an
  etched network: measurement of a fractal dimension. {\it Phys Rev
    Lett}. (1985) {\bf 54}:2226. doi: 10.1103/PhysRevLett.54.2226

\bibitem{lmtsm04} L{\o}voll G, M\'eheust Y, Toussaint R, Schmittbuhl
  J, M{\aa}l{\o}y KJ. Growth activity during fingering in a porous
  Hele-Shaw cell. {\it Phys Rev E}. (2004) {\bf 70}:026301. doi:
  10.1103/PhysRevE.70.026301
  
\bibitem{ww83} Wilkinson D, Willemsen JF. Invasion percolation: a new
  form of percolation theory. {\it J Phys A: Math Gen}. (1983) {\bf
    16}:3365. doi: 10.1088/0305-4470/16/14/028

\bibitem{ws81} Witten Jr TA, Sander LM. Diffusion-limited aggregation,
  a kinetic critical phenomenon. {\it Phys Rev E}. (1981) {\bf
    47}:1400. doi: 10.1103/PhysRevLett.47.1400

\bibitem{v18} Valavanides M. Review of steady-state two-phase flow in
  porous media: independent variables, universal energy efficiency
  map, critical flow conditions, effective characterization of flow
  and pore network. {\it Transp Porous Media}. (2018) {\bf
    123}:45. doi: 10.1007/s11242-018-1026-1

\bibitem{hsbkgv18} Hansen A, Sinha S, Bedeaux D, Kjelstrup S,
  Gjennestad MA, Vassvik M. Relations between seepage velocities in
  immiscible, incompressible two-phase flow in porous media. {\it
    Transp Porous Media}. (2018) {\bf 125}:565. doi:
  10.1007/s11242-018-1139-6

\bibitem{d56} Darcy H. {\it Les fontaines publiques de la ville de
  Dijon: exposition et application des principes {\`a} suivre et des
  formules {\`a} employer dans les questions de distribution
  d'eau}. Paris: V. Dalamont (1856).

\bibitem{w86} Whitaker S. Flow in porous media I: a theoretical
  derivation of Darcy's law. {\it Transp Porous Media}. (1986) {\bf
    1}:3. doi: 10.1007/BF01036523

\bibitem{tkrlmtf09} Tallakstad KT, Knudsen HA, Ramstad T, L{\o}voll G,
  M{\aa}l{\o}y KJ, Toussaint R, Flekk{\o}y EG. Steady-state two-phase
  flow in porous media: statistics and transport properties. {\it Phys
    Rev Lett}. (2009) {\bf 102}:074502. doi:
  10.1103/PhysRevLett.102.074502

\bibitem{tlkrfm09} Tallakstad KT, L{\o}voll G, Knudsen HA, Ramstad T,
  Flekk{\o}y EG, M{\aa}l{\o}y KJ. Steady-state, simultaneous two-phase
  flow in porous media: an experimental study. {\it Phys Rev
    E}. (2009) {\bf 80}:036308. doi: 10.1103/PhysRevE.80.036308

\bibitem{rcs11} Rassi EM, Codd SL, Seymour JD. Nuclear magnetic
  resonance characterization of the stationary dynamics of partially
  saturated media during steady-state infiltration flow. {\it New J
    Phys}. (2011) {\bf 13}:015007. doi: 10.1088/1367-2630/13/1/015007

\bibitem{sbdk17} Sinha S, Bender AT, Danczyk M, Keepseagle K, Prather
  CA, Bray JM, Thrane LW, Seymour JD, Codd SL, Hansen A. Effective
  rheology of two-phase flow in three-dimensional porous media:
  Experiment and simulation. {\it Transp Porous Media}. (2017) {\bf
    119}:77. doi: 10.1007/s11242-017-0874-4

\bibitem{wb36} Wyckoff RD, Botset HG. The flow of gas‐liquid mixtures
  through unconsolidated sands. {\it J Appl Phys}. (1936) {\bf
    7}:325. doi: 10.1063/1.1745402

\bibitem{shbk13} Sinha S, Hansen A, Bedeaux D, Kjelstrup S. Effective
  rheology of bubbles moving in a capillary tube. {\it Phys Rev
    E}. (2013) {\bf 87}:025001. doi: 10.1103/PhysRevE.87.025001

\bibitem{sh12} Sinha S, Hansen A. Effective rheology of immiscible
  two-phase flow in porous media. {\it Europhys Lett}. (2012) {\bf
    99}:44004. doi: 10.1209/0295-5075/99/44004

\bibitem{ct15} Chevalier T, Talon L. Generalization of Darcy's law for
  Bingham fluids in porous media: from flow-field statistics to the
  flow-rate regimes. {\it Phys Rev E}. (2015) {\bf 91}:023011. doi:
  10.1103/PhysRevE.91.023011
  
\bibitem{rh87} Roux S, Herrmann HJ. Disorder-induced nonlinear
  conductivity. {\it Europhys Lett}. (1987) {\bf 4}:1227. doi:
  10.1209/0295-5075/4/11/003

\bibitem{s53} Scheidegger AE. Theoretical models of porous
  matter. {\it Producers Monthly}. (1953) {\bf 17}:17.

\bibitem{s74} Scheidegger AE. {\it The physics of flow through porous
  media}. Toronto: University of Toronto Press (1974).

\bibitem{hhp15} Hansen A, Hemmer PC, Pradhan S. {\it The fiber bundle
  model: Modeling failure in materials}. Berlin: Wiley (2015).

\bibitem{sshb16} Savani I, Sinha S, Hansen A, Bedeaux D, Kjelstrup s,
  Vassvik M. A Monte Carlo algorithm for immiscible two-phase flow in
  porous media. {\it Transp Porous Media}. (2016) {\bf 116}:869. doi:
  10.1007/s11242-016-0804-x


\end{thebibliography}
\end{document}